# Energy dependent Chemical Interface Damping induced by 1-Decanethiol Self-Assembled Monolayer on Au(111)

Maurice Pfeiffer[1], Gregor B. Vonbun-Feldbauer[2,3,4], Jacob B. Khurgin[5], Olga Matts[6], Ahmed Shqer[3], Nadiia Mameka[6], Manfred Eich[1,7], Alexander Petrov*[1,7]

[1] Institute of Optical and Electronic Materials, Hamburg University of Technology, 21073 Hamburg, Germany

[2] Institute for Interface Physics and Engineering, Hamburg University of Technology, 21073 Hamburg, Germany

[3] Institute of Advanced Ceramics, Hamburg University of Technology, 21073 Hamburg, Germany

[4] Institute of Surface Science, Helmholtz-Zentrum Hereon, 21502 Geesthacht, Germany

[5] Department of Electrical and Computer Engineering, Johns Hopkins University, Baltimore, Maryland 21218, United States

[6] Institute of Hydrogen Technology, Helmholtz-Zentrum Hereon, 21502 Geesthacht, Germany

[7] Institute of Functional Materials for Sustainability, Helmholtz-Zentrum Hereon, 14157 Teltow, Germany

*a.petrov@tuhh.de

**Abstract:**

The chemical interface damping (CID) effect increases the collision frequency of free electrons in metals by changes of the metal surface. We have now experimentally disentangled the two contributions to CID: induced roughness and direct charge transfer. The latter is an important area of research in photoelectrochemistry with potential applications in light-induced chemical reactions. We present a broadband investigation of the CID effect on Au(111) covered by a self-assembled monolayer of decanethiol. Spectroscopic ellipsometry measurements show a photon energy dependent increase of collision frequency. We observe a constant, photon energy independent contribution, which is attributed to induced roughness and a contribution that linearly increases with photon energy from about 1 eV upwards which we attribute to direct charge transfer. The onset of the charge transfer mechanism corresponds to occupied orbitals of thiols bound to the Au surface, as confirmed by density functional theory calculations.

**Introduction:**

Charge transfer following light absorption in plasmonic structures is an important area of research in photoelectrochemistry with potential applications in light-induced chemical reactions in general,[1] and more specifically, in water splitting.[2,3] One factor that drives light absorption in metals is the chemical interface damping (CID) effect, which increases the collision frequency (damping rate) of free electrons in metals by altering the metal surface.[4] When properly understood and tailored, this effect may increase the optical absorption in plasmonic media, potentially leading to stronger charge transfer rates. Generally, the optical properties of metals are governed by interband and intraband transitions of electrons. While the interband transitions are direct electronic transitions from narrow energy bands, the intraband transitions are indirect transitions available for a broad spectrum of electrons, which are often approximated as free electrons. The optical response of free electrons in a metal is approximated by the Drude model.[5,6] Within the Drude model the collision frequency describes the rate, thus the probability per unit time of free electrons to undergo intraband transitions. It is called collision frequency as momentum conservation requires a free electron to interact (collide) with an obstacle to assist in the indirect transition. Such obstacles are: phonons, which result in temperature-dependent metal optical properties; lattice defects and grain boundaries, which result in a dependence on the internal structure of the metal; the surfaces/interfaces, which add a dependence on the surface structure and chemical composition. Landau damping[7] describes the roughness independent collision frequency with a metal interface. An additional contribution can originate from initial surface roughness. However, in case of a modified surface, e.g. due to a molecular monolayer, the CID effect adds to the collision frequency.[8,9] The CID contribution is especially important for systems with large surface to volume ratios such as nanoparticles or nanoporous structures, reaching 50% for 5 nm particles.[10] Additionally, collisions between free electrons enable intraband transitions via the Umklapp process.[7] Importantly, the influence of these electron-electron collisions is energy dependent.[11–14]

The CID effect can occur via two different mechanisms.[15] The first mechanism is chemistry induced effective roughness of the surface, which results in additional electron scattering at the surface thereby enabling intraband transitions. The second mechanism is direct charge transfer of an, in case of gold, sp-electron into a surface state or of an electron from a surface state into the sp-band.[15] In the case of thiols, this state is the highest occupied molecular orbital (HOMO) or the lowest unoccupied molecular orbital (LUMO), which are formed due to the attached molecule. We consider hybrid orbitals formed between the thiol and the metal. This mechanism results in a direct transition but still involves a free electron and thus contributes to the effective collision frequency within the Drude model. It should be mentioned that the charge transfer could be also temporal in which case the charge returns back into the metal if it is not participating in a chemical reaction and replaced in a current circuit. Both mechanisms increase the collision frequency. While the roughness-based effect is generally expected to be of broadband nature with contributions even at zero (DC) frequency, thus to electrical resistivity, electronic transitions employing direct charge transfer require a certain photon energy to overcome the potential barrier. It is important to differentiate between both mechanisms when looking at applications such as photocatalysis, where the direct charge transfer mechanism is more beneficial.[15]

Many studies[10,16–20] have investigated the CID effect from thiols, which form a self-assembled monolayer (SAM) on gold.[21–24] This SAM modifies the gold surface and induces CID. Foerster et al.[10] demonstrated the CID effect on gold nanoparticles functionalized with dodecanethiol. In another experiment,[16] they found the CID effect to depend on the thiol molecules. Differences in CID were explained by the influence of the magnitude and direction of the induced surface dipole from the attached adsorbate. This has also recently been confirmed by Stefancu et al.[25] The CID effect also occurs at metal semiconductor interfaces. Several studies[26–28] investigated its role for such interfaces with metal nanoparticles and found a significant increase in the metal collision frequency, which was attributed to direct charge transfer.[29,30]

At the same time the contribution of roughness and charge transfer effects to CID at optical frequencies is still under the debate.[15,25] Almost all previous studies on CID investigated the CID effect on metal nanoparticles. These nanoparticles offer the advantage of a large surface to volume ratio, resulting in a strong response from surface modifications. However, the effect of CID was evaluated by the change in the width of the plasmonic resonances limiting the investigation to discrete wavelengths. Prato et al.[31] investigated the optical response of different alkanethiol coatings on gold with spectroscopic ellipsometry, but their considerations were limited to the visible spectrum. They concluded that only a change in the Drude electron collisions, in fact CID, can explain the observed behavior for red light. We recently investigated the CID effect from the electrochemical oxidation of gold using spectroscopic ellipsometry,[32] but the spectral width was also limited due to IR absorption of water. Two studies[33,34] investigated the change of the DC resistivity of copper as well its IR reflectivity upon adsorption of CO and $C_2H_4$ and found a direct link between the reflectivity and the resistivity, explaining the observed changes with increased scattering of electrons due to the additional scattering centers formed by the adsorbates. Stefancu et al.[25] recently conducted a study of CID from different molecules on gold in the DC and at two wavelengths in IR range, concluding on an increasing CID effect at larger photon energies for biphenyl thiol due to the charge transfer mechanism. Recently, Yuan et al.[35] managed to extract the energy dependence of CID from measurements of single particles covered by $TiO_2$ and attributed it to the charge transfer mechanism as well.

In this work we present a broadband study that aims to distinguish the different mechanisms of the CID effect. We use template-stripped Au(111) substrates and coat them with a decanethiol SAM to obtain a simple yet effective system showing CID. The optical properties of the bare and coated samples are investigated with spectroscopic ellipsometry for photon energies between 0.5 eV (2500nm) and 1.75 eV (708 nm). Importantly, we apply a pointwise fitting procedure that provides the collision frequency at each wavelength independently. This approach allows us to look at the spectral behavior of CID without prior assumptions. The spectral position for the onset of the direct charge transfer contribution is theoretically assessed via the projected density of states (pDOS) of gold atoms at the surface, which are bound to thiols, using density functional theory (DFT).

## Methods:

Gold Samples: We used single-crystal template-stripped gold nanolayers (100 nm thickness) on glass chips purchased from Platypus Technologies. These layers have an RMS roughness of 0.36 nm and show a primarily Au(111) structure with 92% of grains, as measured by the manufacturer. Thus, we will refer to them as Au(111) layers.

Thiol Coating: The samples were immersed in a 3mM decanethiol ethanolic solution for 4 hours, following a procedure used in Ref.[10], to grow a thiol SAM on the Au(111) surface. After the immersion they were carefully washed by shortly submerging first in pure ethanol and afterwards in pure isopropanol.

Optical Response Model: The optical properties of bulk gold below 1.55 eV (above 800 nm) are well described by the Drude model $\varepsilon_\text{D}(\omega) = \varepsilon_\infty - \frac{\omega_\text{p}^2}{\omega^2 + i\gamma\omega}$ with $\varepsilon_\infty$ being used to account for contribution from all electronic transitions at higher frequencies, the plasma frequency $\omega_\text{p}$ and the collision frequency $\gamma$ which is considered as a constant in the original model.[5,6] Modern versions of this theory include an energy dependent collision frequency which follows $\gamma(\omega) = \gamma_0 + \beta_\text{ee}(\hbar\omega)^2$ with $\beta_\text{ee}$ describing the energy dependent electron-electron scattering contribution.[11–14] Further, at energies above 1.55 eV we observe a slight deviation of the gold properties from the Drude behavior including the e-e-scattering term. This deviation is caused by the onset of electronic interband transitions from the d-band to the Fermi edge in the sp-band. For proper modelling of this interband transition, we include a Lorentz oscillator term. It follows $\varepsilon_\text{L}(\omega) = \frac{f}{\omega_\text{L}^2 - \omega^2 - i\Gamma_\text{L}\omega}$ with the oscillator strength f, the resonance position of the oscillator $\omega_\text{L}$ and the resonance broadening parameter $\Gamma_\text{L}$.[36] At higher energies the contribution of the interband transitions becomes more significant. We avoid this spectral region to prevent the masking of the observed CID effect. It should be mentioned that the excitation of gold by an external plane wave does not allow to obtain significant electric fields orthogonal to the surface even at oblique angles of incidence ($|E_\perp|/|E_\parallel| = |k_\parallel|/|k_\perp| \ll 1$, where $E$ and $k$ are the electric field and wavenumber inside the metal, correspondingly). Thus, the Landau damping contribution, which requires an electric field normal to the surface, is assumed to be negligible in our results. Further, we do not specify a model for the energy dependence of the collision frequency change due to the CID effect, but use pointwise fitting, which is detailed in the following section.

Ellipsometry Measurements and Fitting: The optical properties of the gold samples were measured with a spectroscopic ellipsometer (J. A. Woollam RC2) at five different angles of the incident light (55°, 60°, 65°, 70° and 75°). In this work we employ two different fitting techniques to our ellipsometry measurements – spectral and pointwise fitting.

The spectral fitting involves simultaneous fitting over a specified spectral range and requires the assumption of a model such as described in the "Optical Response Model" section above. We fit $\omega_p, \gamma_0, \beta_{ee}, f, \omega_L, \Gamma_L$ to the measurement results obtained on Au(111) prior to coating with thiols. Note that e-e-scattering and the onset of interband transitions have different energy dependences and thus can be safely fitted simultaneously. For the case with a thiol monolayer on top, we employ a spectral fit only to obtain the effective layer thickness of the thiol monolayer. Here, we assume the additional collision frequency $\gamma_{CID}$ from CID to be spectrally constant. To justify this assumption, we only conduct this fit over a small spectral range from 1000 to 1050 nm. During this fit the previously obtained parameters from the pure gold fit are used as constants.

With pointwise fitting we aim to fit the collision frequency for each spectral energy without prior assumption of a model for CID. We conduct the pointwise fitting for both cases using $\omega_p, f, \omega_L, \Gamma_L$ obtained from spectral fitting and fit the collision frequency $\gamma$ for every energy point. In the case with thiols, we include the thiol layer with the prior fitted thickness. The extent of CID can then be extracted from the pointwise fitting results via $\gamma_{CID}(\hbar\omega) = \gamma_{Au,thiol}(\hbar\omega) - \gamma_{Au}(\hbar\omega)$.

Density functional theory (DFT) calculations: Spin-polarized DFT calculations were done using the Perdew-Burke-Ernzerhof (PBE) exchange correlation functional[37] on atomic sulfur (S), thiol groups (SH), and methanethiols ($SCH_3$) adsorbed on ideal Au(111) slabs and on Au adatom covered Au(111) slabs, which serve as simplified model systems. Formally in the adsorption of methanethiols ($HSCH_3$) on Au(111), thiyl radicals ($SCH_3$) and $H_2$ are formed, but in the relevant literature the $SCH_3$ is often referred to as thiol or thiolate and we use the naming thiol here as well. The DFT calculations were performed with the Vienna Ab Initio Simulation Package (VASP, version 5.4.4[38–41]). The computational setup follows previous work on OH adsorption on Au(111)[2] and details can be found in the supporting information S5. An adsorbate coverage of 1 per 4 Au atoms at the surface was assumed if not stated otherwise. At this coverage, S, SH, and $SCH_3$ were initially placed at fcc adsorption sites considering only a single adsorbate geometry per model system. Adsorbates starting in an upright geometry maintained their geometry and site during relaxations. In models with initially tilted $SCH_3$, they, however, relaxed into a bridge-like site. Additionally, surfaces with Au adatoms and more complex adsorbate overlayers and coverages of 1 per 3 and 3 per 8 Au atoms at the surface were used for comparison and taken from the work of Mom et al.[42]

**Results and Discussion:**

The fresh gold samples were optically measured with the spectroscopic ellipsometer. Directly after this measurement the samples were immersed for 4h in decanethiol solution followed by careful washing in ethanol and isopropanol. During this time a SAM of decanethiol arranged on the gold surface. Following the coating process the samples were measured again with the ellipsometer to identify the changes to the optical properties of the gold induced by the monolayer of decanethiol. A schematic of the experimental setup is shown in Supporting Information S1. The immersion time of 4h was chosen based on the experiments by Foerster et al.[10] They investigated gold nanoparticles functionalized by similar 1-dodecanethiol and found the CID effect to saturate after about 1h also using 3mM solution. We quadrupled this time to make sure we have a saturated monolayer of decanethiol on our gold samples.

The ellipsometric fitting of the thiol monolayer on top of the gold substrate resulted in a layer thickness of 1.56 nm and 1.63 nm for the two samples. This fits well with theoretical predictions of 1.5 nm[21] and experimental findings based on angle-resolved X-ray photoelectron spectroscopy[24] that show a thickness of 1.51 nm. For our fits we assumed a refractive index of 1.451 based on the liquid phase of the decanethiol.[43]

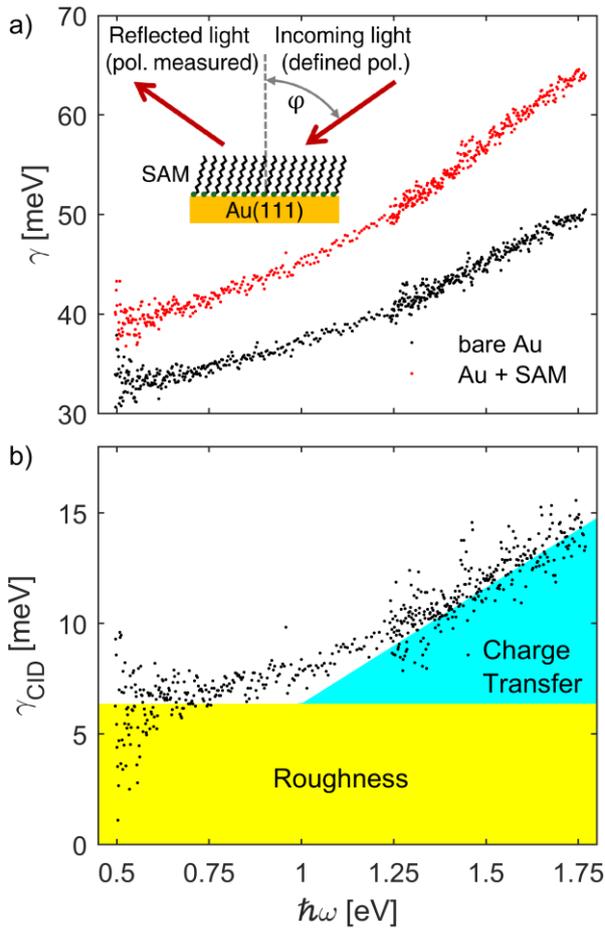

***Figure 1:*** *Pointwise fitted collision frequency and extent of CID effect vs. photon energy. The values represent the mean over 5 measured angles. a) Collision frequency of Au(111) prior to thiol coating as well as after decanethiol coating. The increase of Au(111) collision frequency at higher energies originates from the photon energy dependent electron-electron scattering contribution (see Supporting Information S3). An overall increase of the collision frequency from the coating is observed. The inset shows a schematic of the experiment (see Supporting Information S1). b) Collision frequency change due to the CID effect calculated by subtracting the values in a). We observe an increase of the CID effect towards higher energies. The jump in noise at about 1.25 eV originates from a detector change. The noise at low energies originates from low intensity of the ellipsometer light source at this end of the spectral range. The colored areas indicate the mechanisms of CID.*

The fitting of the ellipsometry measurements resulted in a pointwise collision frequency determination as depicted in Figure 1. Additional data from a second sample showing good overlap with the results presented here is shown in Supporting Information S2. The black dots in Figure 1a show the collision frequency of the uncoated Au(111) sample. We observe an increase of collision frequency towards higher energies caused by electron-electron scattering. Spectral fitting with the Drude model as detailed in the methods section resulted in the electron-electron scattering related collision frequency coefficient $\beta_{ee} = 0.0059\ eV^{-1}$ (see more details in Supporting Information S3) and the energy independent collision frequency $\gamma_0 = 32\ meV$ (20 fs decay time) which both fit well to other measurements.[25,44–48] The energy independence of the roughness contribution suggests that the spatial spectrum of the roughness is essentially flat. This is, of course, an approximation — in reality, the roughness contribution is expected to decrease at higher photon energies. However, such a decrease appears to occur only at energies higher than the maximum energy explored in our experiments (1.75 eV).

The red dots in Figure 1a show the collision frequency of the Au(111) with decanethiol monolayer on top. We observe an overall elevated collision frequency due to CID. The absolute increase was calculated by subtracting the bare Au(111) results from the results with Au(111) coated with decanethiol. It is depicted in Figure 1b and shows the extent of CID induced by the surface functionalization. At low energy we observe a collision frequency change due to CID of approx. 6.4 meV (averaged from 0.5 to 0.75 eV) which is largely constant with respect to energy. At higher energies the CID contribution gradually increases. We attribute the low energy collision frequency change to the roughness effect that should be present even in DC measurements and the increase of CID at higher photon energies to the charge transfer contribution.[15] With this consideration we can estimate that direct charge transfer contributes more than 50% to the CID effect at 1.75 eV.

In Supporting Information S4 we provide a comparison of our findings to DC resistivity measurements of gold nanostructures and surfaces covered with thiols. A comparison of the total CID effect at optical frequencies including the charge transfer contribution can be done with a study of Foerster et al.[10] They measured the increase in the collision frequency of gold nanoparticles from 1-dodecanethiol coating at plasmonic resonance of about 1.7 eV. Scaling the effective mean free path of 11 to 25 nm of their nanoparticles to 52 nm of the flat gold[32] as described in section S4, they would observe an increase of collision frequency of about 10 meV on flat gold which matches reasonably well with our 14 meV at 1.7 eV photon energy. We argue that we likely observe a larger increase due to the initial low roughness of our 111-oriented single-crystalline samples. A study by Stefancu et al.[25] recently reported a similar behavior when coating Au with thiols. For 1-dodecanethiol they measured CID of 5 meV at 1.57 eV, which is smaller than our results. For biphenyl thiol they investigated CID at two energies 1.44 eV and 1.57 eV and observed a stronger effect at the larger energy, which is attributed to increasing charge transfer contribution. Note that they consider the charge transfer into LUMO for biphenyl thiol. For 1-decanethiol the LUMO is not reachable with the photon energies used and we concentrate on the electron excitation from HOMO into states slightly above the Fermi level as the charge transfer mechanism.

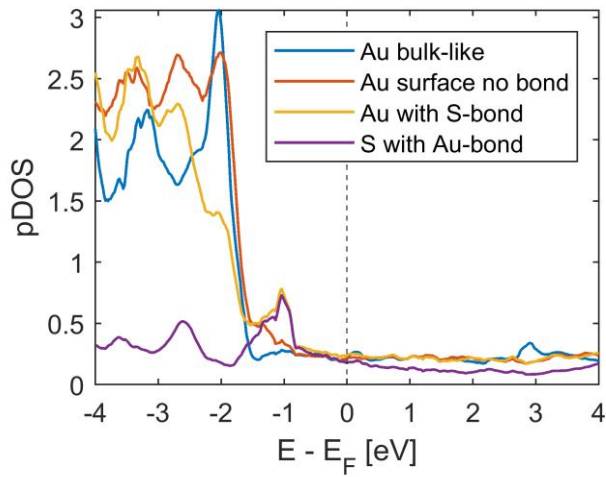

*Figure 2:* Projected density of states relative to Fermi energy for different selected atoms of Au(111) with $SCH_3$ adsorbates at bridge adsorption sites with 0.25 monolayer coverage relative to the Au surface. The corresponding structure is shown in supporting information S5. In pure gold (blue and red curves) the presence of the d-band is seen as an increase in DOS at about 2 eV below the Fermi level, whereas the formation of covalent bonds between gold and sulfur leads to an increase of the DOS at around -1 eV which is where the hybrid HOMO states of the adsorbate start to contribute (yellow and purple curves).

From previously presented theory we expect an energy independent contribution due to roughness and a stepwise increase in collision frequency when charge transfer becomes possible, which is followed by an increase proportional to photon energy.[15] Figure 2 shows the pDOS for single atoms in a model system corresponding to Au(111) with methanethiol adsorbates. Other possible model systems are discussed in the supporting information S5 and S6 showing similar results. It shows that the hybrid HOMO from the bound thiol lies around 1.0 eV below the Fermi energy. Thus, we expect the charge transfer mechanism of CID to have an onset around this energy and to grow in the range between approx. 1.0 and 1.5 eV with following saturation. Figure 1 shows that we observe a gradual increase in collision frequency starting at around this energy, but no saturation above 1.5 eV as expected from the integral over the DOS. At the same time the DOS alone does not define the transition probability and more detailed quantum mechanical analysis of electronic states is required. Also, in the former estimation of the CID effect[15] via direct charge transfer we assumed that the charge leaves the metal. At the same time in our thiol system the electron directly excited from HOMO returns to the adsorbate after some time. If the temporal transfer times are smaller than the plasmon decay time, then our estimation has to be modified. Recently, an ab-initio estimation was conducted for thiols using DFT calculations and theory developed by Persson which also predict growing CID contribution at larger photon energies.[25,49,50]

## Conclusion:

To conclude, we report the first broadband spectral investigation of the CID effect with a model system of Au(111) coated with a decanethiol monolayer. We observe a spectral dependence of the CID effect with a convergence to a constant contribution at low energy and a linear increasing contribution at photon energies above the potential barrier for excitation of electrons from the HOMO of the surface-bound thiol. We attribute the constant contribution to roughness effect induced by thiol molecules and the linear rise above 1 eV to charge transfer effect. At 1.75 eV more than 50% of electrons excited by CID are directly transferred into adsorbate states.

We show that spectroscopic ellipsometry can be used to reliably determine the changes in collision frequency due to surface effects such as CID. Importantly, the point-by-point fitting approach allows us to determine the effect at each frequency independently, thereby avoiding the biases that can arise when assuming specific dispersion models. The obtained spectral dependency of CID helps to better understand the charge transfer efficiencies in nanostructures. Further work will be concentrated on tests of other molecules and dielectric coatings on metal surface.

## TOC/Abstract Graphics:

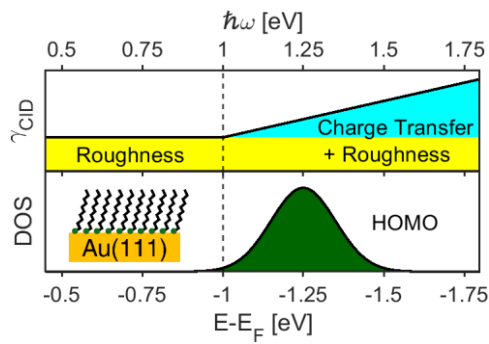

## Acknowledgement:

We thank Mauricio Schieda and Francesca Toma from Helmholtz Zentrum Hereon for providing access and introduction to their spectroscopic ellipsometer.

This work was supported by Deutsche Forschungsgemeinschaft (DFG) through Project 192346071, SFB 986.

Supporting Information for

# Energy dependent Chemical Interface Damping induced by 1-Decanethiol Self-Assembled Monolayer on Au(111)

Maurice Pfeiffer[1], Gregor Vonbun-Feldbauer[2,3,4], Jacob Khurgin[5], Olga Matts[6], Ahmed Shqer[3], Nadiia Mameka[6], Manfred Eich[1,7], Alexander Petrov*[1,7]

[1] Institute of Optical and Electronic Materials, Hamburg University of Technology, 21073 Hamburg, Germany

[2] Institute for Interface Physics and Engineering, Hamburg University of Technology, 21073 Hamburg, Germany

[3] Institute of Advanced Ceramics, Hamburg University of Technology, 21073 Hamburg, Germany

[4] Institute of Surface Science, Helmholtz-Zentrum Hereon, 21502 Geesthacht, Germany

[5] Department of Electrical and Computer Engineering, Johns Hopkins University, Baltimore, Maryland 21218, United States

[6] Institute of Hydrogen Technology, Helmholtz-Zentrum Hereon, 21502 Geesthacht, Germany

[7] Institute of Functional Materials for Sustainability, Helmholtz-Zentrum Hereon, 14157 Teltow, Germany

*a.petrov@tuhh.de

## S1: Experimental Setup

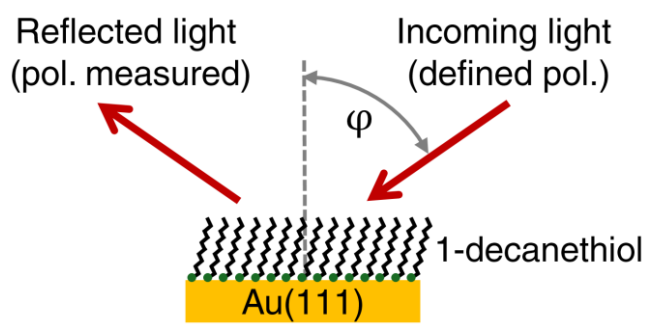

*Figure SI-1:* Schematic of the experimental setup. IR and visible polarized light is directed onto the sample under specified incidence angle φ. Upon reflection the light changes its polarization, which is measured by the ellipsometer. Here we depict the case with a decanethiol monolayer attached to the Au(111) surface.

Figure SI-1 shows a simplified schematic of the experimental setup. A parallel beam with defined polarization is reflected from the gold surface under a specified incidence angle φ. Upon reflection, the polarization of the light changes depending on the optical properties of the Au(111) sample and if present, on the properties of the decanethiol monolayer. The polarization of the reflected light is measured and the change of polarization calculated and represented by the Ψ (amplitude ratio) and Δ (phase difference) parameters. These parameters are then used within fitting procedures to obtain information about the optical properties of the material stack under investigation. The fitting procedure and the used optical model is discussed in the methods section. With respect to the molecular tilt angle and the associated azimuthal orientation of the molecules shown in Figure SI-1 we assume domain formation on the surface with the area illuminated from the incoming beam being much larger than a single domain. Thus, in our measurement setup we average over all possible azimuthal domain orientations.

## S2: Collision frequency of two samples

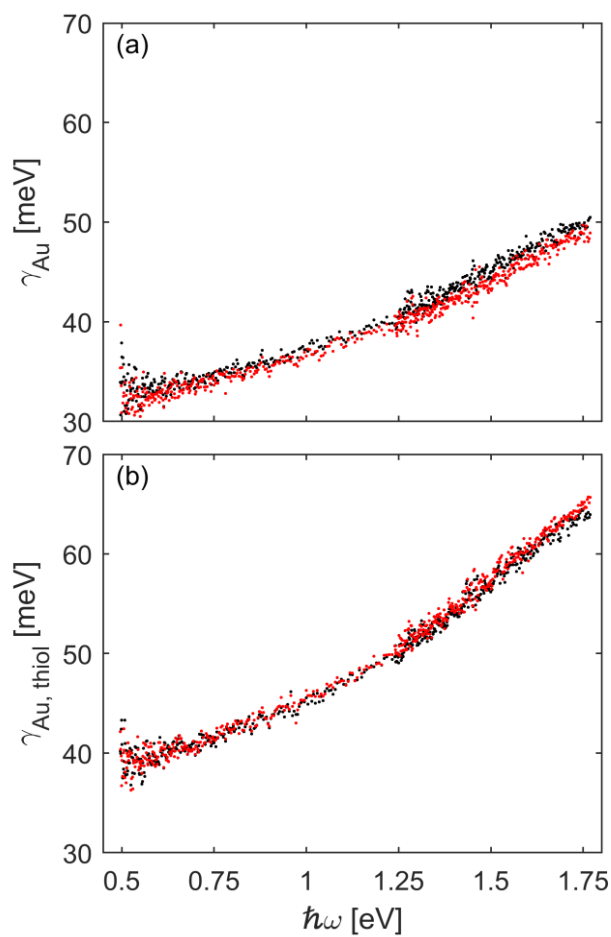

***Figure SI-2:*** *Fitted collision frequency vs. photon energy for two samples. The values represent the mean from averaging over the fitting results from all measured angles. The black dots are the results from the same sample as discussed in the main text and the red dots are the results from a second sample. a) Collision frequency of Au(111) prior to thiol coating. b) Collision frequency of Au(111) coated with decanethiol SAM. We observe a good overlap between both samples.*

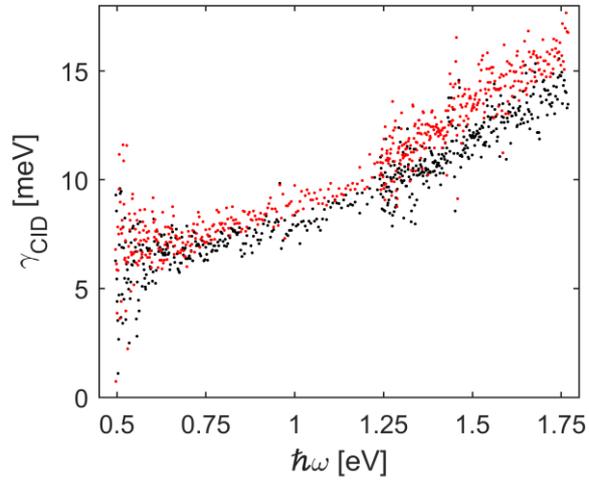

***Figure SI-3:*** *Extent of CID effect vs. photon energy for two samples: black dots indicate the sample discussed in the main text and red dots the second sample. We observe a quite good overlap between both samples.*

## S3: Consideration of energy dependent electron-electron scattering collision frequency

Many studies[51–59] assumed a constant collision frequency when applying the Drude model to gold or metals in general. This disregards the electron-electron scattering contribution to the collision frequency, which scales quadratically with frequency.[11–14] For narrowband or IR investigations one may disregard this, but especially in the visible light region the consideration of energy dependent electron-electron scattering is important to draw precise conclusions when spectrally applying the Drude model.

As we carefully look at the collision frequency, we need to make sure our fitting parameters are accurately determined. Thus, the consideration of the energy dependent contribution of electron-electron collisions is necessary.[11–14] Within the Drude model we use the photon energy dependent collision frequency $\gamma(\omega) = \gamma_0 + \beta_{ee}(\hbar\omega)^2$ to account for this effect. The energy independent part of the electron-electron collision frequency is included in the $\gamma_0 = 32$ meV together with other energy independent contributions such as electron-phonon, electron-grain boundary scattering and initial-roughness scattering. It fits well with other measurements, which is discussed in the Supporting Information S4.

According to Parkins et al.[13] the electron-electron scattering contribution to the overall collision frequency follows $\gamma_{ee} = \frac{1}{12}\pi^3 \Gamma\Delta \frac{1}{\hbar E_F}\left[(k_B T_e)^2 + \left(\frac{\hbar\omega}{2\pi}\right)^2\right]$. It has a temperature and an energy dependent contribution. Within our considerations we do not explicitly look at the temperature dependent $(k_B T_e)^2$ term. This contribution is small compared to energy dependent term and is included within the constant $\gamma_0$, as our experiments are conducted at constant room temperature. We assume the light intensity of the ellipsometer to be sufficiently low to assume the electron temperature to be close to the lattice temperature. The energy dependent part alone follows $\gamma_{ee} = \frac{\pi\Gamma\Delta}{48}\frac{1}{\hbar E_F}(\hbar\omega)^2 = \beta_{ee}(\hbar\omega)^2$ with the average scattering probability $\Gamma = 0.55$, the fractional Umklapp scattering $\Delta = 0.75$ and the Fermi energy $E_F = 5.53$ eV for gold.[12] It results in $\beta_{ee} = 0.0049$ eV$^{-1}$, which matches reasonably well with our experimental results presented in the main text. An experiment by Cao et al. found electron-electron scattering to result in a lifetime of 85 fs at 1 eV ($\beta_{ee} = 0.0077$ eV$^{-1}$),[44] which is close to our findings.

## S4: Literature comparison of measured CID effect at optical frequencies and DC

Most optical studies of gold (e.g.[51–59]) fit the Drude model with a constant collision frequency $\gamma_0$ and thus lack consideration of the photon energy dependent electron-electron collision frequency contribution that was discussed in S3. Consequently, the reported $\gamma_0$ appears larger than its actual value, as it includes the electron-electron scattering contribution at a specific energy or averaged over an energy spectrum. Therefore, these studies are not suitable for comparison with our $\gamma_0$ of 32 meV. But there are also studies investigating the specific resistance of gold $\rho_{Au}$, which is directly linked to the energy independent part of the electron collision frequency $\gamma_0$ via $\rho_{Au} = \frac{m_e \gamma_0}{n_{Au} e^2}$.[60] In literature $\rho_{Au}$ is reported in many studies. Here we present results from four publications stating $\rho_{Au}$ to be 2.2 µΩ cm,[45,46] around 2.3 µΩ cm[47] or 2.44 µΩ cm.[48] With the electron mass $m_e$, the free electron density of gold $n_{Au} = 5.9 \cdot 10^{28}\ m^{-3}$,[60] the elementary charge $e$ and the specific resistance of 2.44 µΩ cm, we calculate the energy independent collision frequency 27 meV. This value is slightly smaller compared to the 32 meV we obtain from spectral ellipsometric fitting. This deviation may be explained by the sensitivity of ellipsometry to the surface, which adds electron scattering at the film surface to the fitted value.[14] But also, differences in the structure and morphology of the samples may explain it. The recent study by Stefancu et al.[25] shows DC resistivity of flat Au films to be about 3 µΩ cm, which corresponds to 33 meV collision frequency and present a very good match with our fitting result of 32 meV.

Similar to the bare Au(111) we compare the observed photon energy independent CID effect of 6.4 meV to CID measured with specific resistance measurements on thiol coated nanostructures. As CID effect depends on the probability of electron-surface collision we need to normalize the results by the effective mean free path of excited electrons between two collisions with the surface $l_{\text{eff}}$.[10] It can be determined from the volume to surface ratio using the mean chord theorem $l_{\text{eff}} = 4V/S$.[61] For our 100 nm thick gold layer the volume is defined not by the thickness of the layer but by the penetration depth of the evanescent optical fields of 13 nm resulting in $l_{\text{eff}} = 52$ nm.[32]

Zhang et al.[62] observed an increase of specific resistance of about 3% from adsorption of similar hexadecanethiol to the 50 nm gold film. They do not provide absolute values of resistivity, making the direct determination of the absolute collision frequency increase not possible. Henriquez et al.[63] investigated the resistivity increase of gold thin films with 10 nm thickness from surface modification with dodecanethiol. They observe a resistivity increase of up to 12%, from which we calculate an increase of collision frequency of about 6 meV based on the absolute resistivity values provided. Furthermore, they find a strong dependence of the observed increase on the surface roughness of the sample, with samples having initial low roughness exhibiting a higher increase in collision frequency by CID. This goes along with the picture that the collision frequency contribution of surface scattering of a flat gold thin film is low and can be significantly increased by surface modifications. For a 10 nm thick layer measured at DC one can assume the exciting electric fields to be constant within the material leading to $l_{eff} = 40$ nm.[61] Thus, they observe an increase in collision frequency which is equivalent to 4.6 meV when scaled to our system ($l_{eff} = 52$ nm[32]). This is similar to the 6.4 meV we observe with our optical investigation. We argue that the significantly lower roughness (0.38 nm) of our template-stripped Au(111) thin films, compared to those used by Henriquez et al. (1.6 nm), may explain the stronger effect observed by us.

## S5: Computational details of Density functional theory calculations

Periodic boundary conditions, Projector Augmented-Wave (PAW) pseudopotentials,[64,65] and the Perdew-Burke-Ernzerhof (PBE) exchange correlation functional[37] were used in the DFT calculations. A plane-wave energy cut-off of 550 eV and a k-point grid of 15 x 15 x 1 for a 2x2 surface cell were chosen allowing for a convergence within 1 meV/atom. The k-point grid was scaled accordingly for larger simulation cells. All structures were relaxed until the forces converged to less than 5 meV/Å. Densities of states (DOS) were calculated using the tetrahedron method with Blöchl corrections.[66] Bader charges were obtained with tools from the Henkelman group.[67–69]

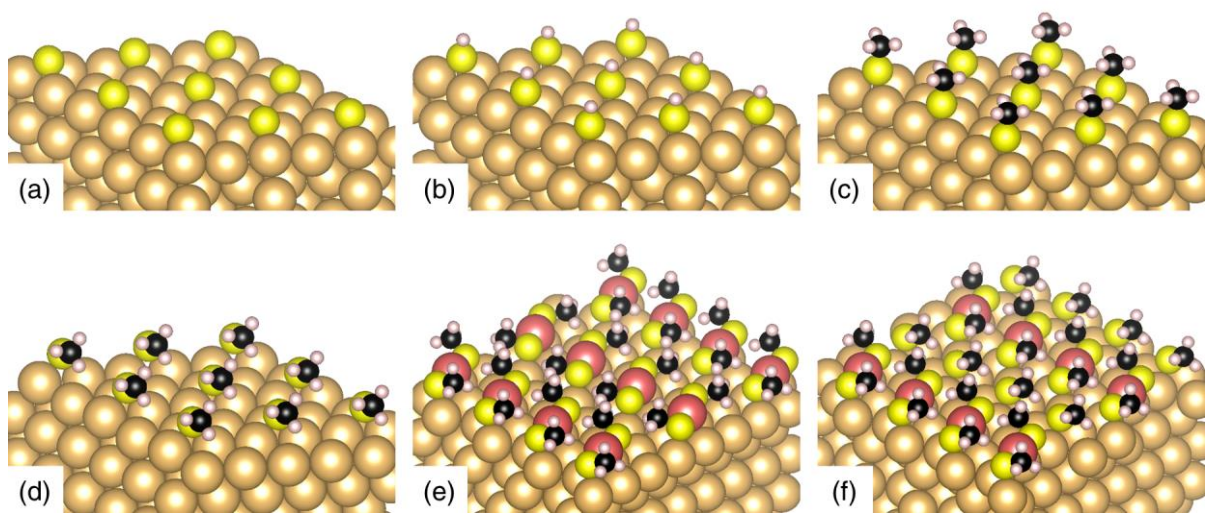

*Figure SI-4:* Adsorbate geometries on Au(111) surfaces used in the DFT calculations. In a) and b) S and SH, respectively, at fcc adsorption sites with a coverage of 1/4 ML are depicted. In c) and d) $SCH_3$ at fcc and bridge-like adsorption sites, respectively, with 1/4 ML coverage are shown. In e) and f) $SCH_3$-Au adatom adsorption structures with thiol coverages of 1/3 ML and 3/8 ML, respectively, are presented. Element color code: Au – pale gold, Au adatom – pastel red, S – yellow, C – black, H – light red.

For the 1/4 ML (where a full monolayer (ML) is defined as one adsorbate per Au atom in the surface layer) adsorbate coverage structures, the surfaces were modeled using periodic 2x2 surface cells and slabs of 12 Au layers separated by a vacuum region of at least 13 Å. Twelve Au-layers were used to obtain a bulk-like DOS for the central Au layers and the adsorbates are placed symmetrically on both sides of the slabs. The relaxed adsorption structures are shown in Figures SI-4a-d.

Additionally, two structures from the work of Mom et al.[42] were taken to represent low ($10^{-6}$ mbar) and high (1 bar) pressure phases with Au adatoms on the surface. The adatoms are assumed to be a result of the adsorption process. For low pressures, the chosen "tc" structure (following the nomenclature in that paper) exhibits $6 \times \sqrt{3}$ surface unit cell and thiol and adatom coverages of 1/3 ML and 1/6 ML, respectively. For high pressures, the "iso1" structure was used, which has a $4 \times \sqrt{3}$ surface unit cell and thiol and adatom coverages of 3/8 ML and 1/8 ML, respectively. For the simulations the unit cells of both structures were doubled along the short lattice vector. For those structures, asymmetric slabs with only five Au layers were used as taken from the literature to minimize the computational costs. The adsorption structures are shown in Figures SI-4e,f.

## S6: Results of DFT calculations for additional model systems

Potential limitations of the model systems used for the computational work are that smaller molecules are used to keep the computational cost low, that the solution, in which the coating process is performed, is not simulated, and that the exact atomistic structure of the Au-thiol interface in the experiments is not known. Several adsorbates and surface models were studied to assess their influence on the DOSs of those models.

S, SH and SCH$_3$ were adsorbed at fcc sites with a coverage of 1/4 ML and the S rest-group bond being perpendicular to the surface to test the influence of rest groups bond to the sulfur adsorbed to Au surfaces. During relaxation those adsorbates stayed at the fcc site with the rest groups pointing away from the surface (see Figures SI-4a-c). A comparison of selected projected DOSs (pDOS) for those adsorbates is shown in Figure SI-5. All three adsorbates show similar features with the hybrid HOMO state around -1 eV where for S, the peak is closest to the Fermi energy. The results are also very similar to previous work on OH adsorption on Au(111),[2] which is not surprising since S is directly below O in the same column of the periodic table.

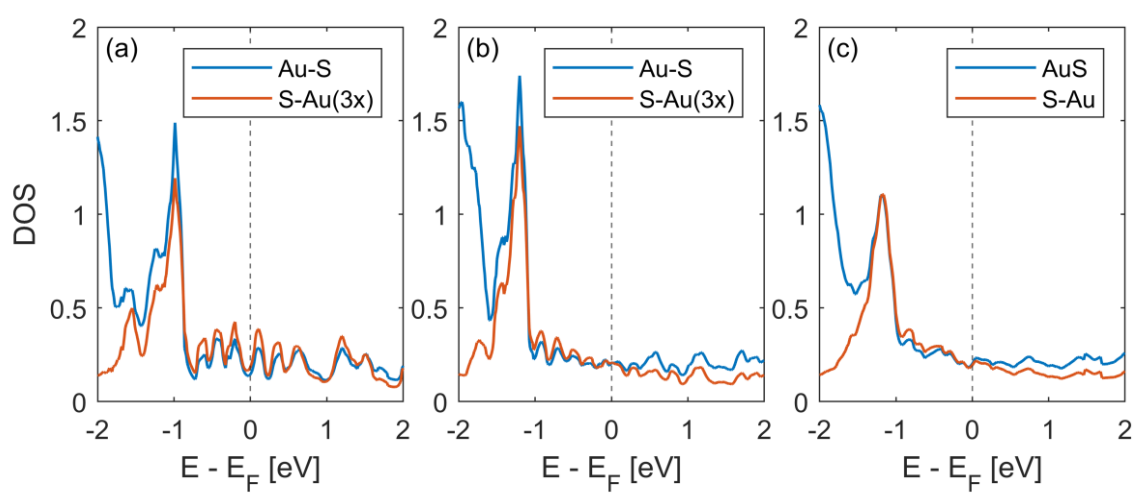

*Figure SI-5: Projected density of states relative to Fermi energy for the adsorbates, a) S, b) SH, and c) SCH$_3$, at fcc adsorption sites with 0.25 monolayer coverage on Au(111). Only the DOSs projected on one S atoms for each adsorbate are shown.*

Additionally, SCH$_3$ was absorbed at the fcc site but the thiol was tilted away from the surface normal, as it is usually observed for thiol SAMs at Au(111) with longer rest group carbon chains.[70] During relaxation, those adsorbates moved to a bridge-like site slightly of the perfect bridge site, shifted about 0.2 Å towards the hcp site (see Figure SI-4d). Adsorption at the bridge-like site leads to a corrugation of the Au(111) surface of about 0.3 Å (measured as the vertical distance between the highest and lowest Au atom within the surface layer) which is increased compared to the fcc structure with 0.18 Å. The bridge-like configuration is energetically more favorable in our calculations by about 0.36 eV/adsorbate compared to an upright methanethiol at an fcc site. According to a Bader charge analysis, the adsorbed thiol S atoms are slightly negatively charged. These S atoms obtain about 0.05 $e$ from each nearest neighbor Au and from the methyl group. This results in a small negative net charge at the S atoms of 0.2 e and 0.15 e for adsorption at an fcc and a bridge site, respectively. These small charge transfers can be rationalized from the small differences and the hierarchy in the Pauling electronegativities of the involved elements (Au: 2.54; C: 2.55; S: 2.58). The redistribution of charge upon adsorption is visualized in Figure SI-6 where the difference of the charge density of the adsorbate structure and the isolated surface and the isolated molecule is shown. A small charge depletion occurs at the surface Au atoms with S bonds in the direction of that bond. For the thiol S, the charge is accumulated roughly pointing towards the surface Au atoms while some charge is depleted in other directions. Because of the asymmetric adsorption geometry, a splitting of peaks in the DOS is expected. Selected pDOSs for fcc and bridge-like adsorption sites are presented in Figure SI-7. For the bridge-like site, the hybrid HOMO state is slightly closer to the Fermi energy and additional peaks are observed towards more negative energies.

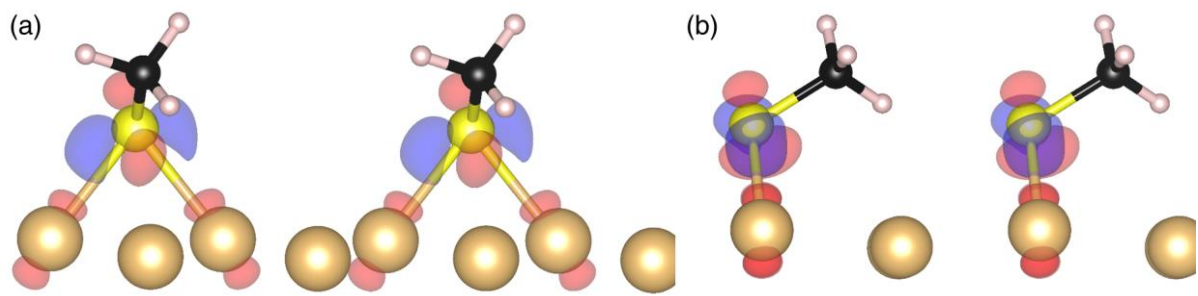

*Figure SI-6: Charge density difference for SCH$_3$ adsorbed on Au(111) at the bridge-like site and the isolated surface and the isolated molecule shown for two rotations of the system. In (a) with the Au-S connection line being roughly parallel to the screen and in (b) with the S-C bond being roughly parallel to the screen. The iso-surface value is 0.01. The red-colored areas represent charge depletion, while the blue-colored areas indicate charge accumulation. Element color code: Au – pale gold, S – yellow, C – black, H – light red.*

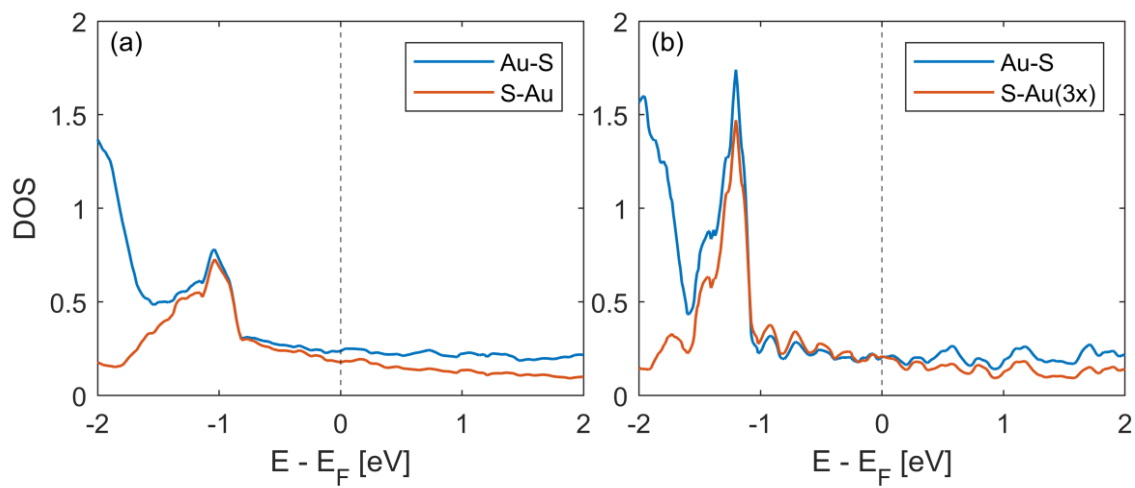

*Figure SI-7:* Projected density of states relative to Fermi energy for methanethiol adsorbates, $SCH_3$, at a) bridge-like and b) fcc adsorption sites with 0.25 monolayer coverage on Au(111). Only the DOSs projected on one S atoms and one Au atom with an Au-S bond for each absorption site are shown.

Moreover, two structures with increased adsorbate coverages of 1/3 and 3/8 ML and Au adatoms on the surface were included as they represent more realistic structures at Au(111)-vacuum/gas interfaces (see S5). The 1/3 ML structure (see Figure SI-4e) is typical for vacuum conditions ($10^{-6}$ mbar) and shows an Au adatom coverage of 1/6 ML. All methanethiols sit almost on top of a surface Au and are additionally bonded to one Au adatom. Those adatoms are located at bridge sites of the Au(111) surface and they are roughly at the same height as the thiol S atoms. Each adatom bonds to two thiols forming 1D rows of thiol-Au-thiol motifs (referred to as "staples" in the literature[42]). The 3/8 ML structure (see Figure SI-4f) was suggest by Mom et al.[42] as a high pressure (1 bar) phase. This phase features a lower adatom coverage of 1/8 ML. Again, rows of thiol-Au-thiol staples are formed; however, involving only 2/3 of the adsorbed thiols in the high-pressure phase. Between two of those rows, there is one parallel row of methanethiols bonded to the surface without an adatom bond. The adsorption geometry of the thiols in these rows closely resembles the bridge-like adsorbates discussed above at 1/4 ML. Selected pDOSs for the two adatom structures and the 1/4 ML bridge-like adsorption structure are presented in Figure SI-8. The hybrid electronic states around -1 eV to the Fermi energy are significantly weaker for the S atoms bond to Au adatoms while the first stronger contribution occurs only around -2 eV. Thus, this scenario does not fit to the experimental results well. The bridge-like adsorbates in the 3/8 ML structure, however, show a very similar behavior to the 1/4 ML case. This indicates that such a structure could also occur at the experimental conditions.

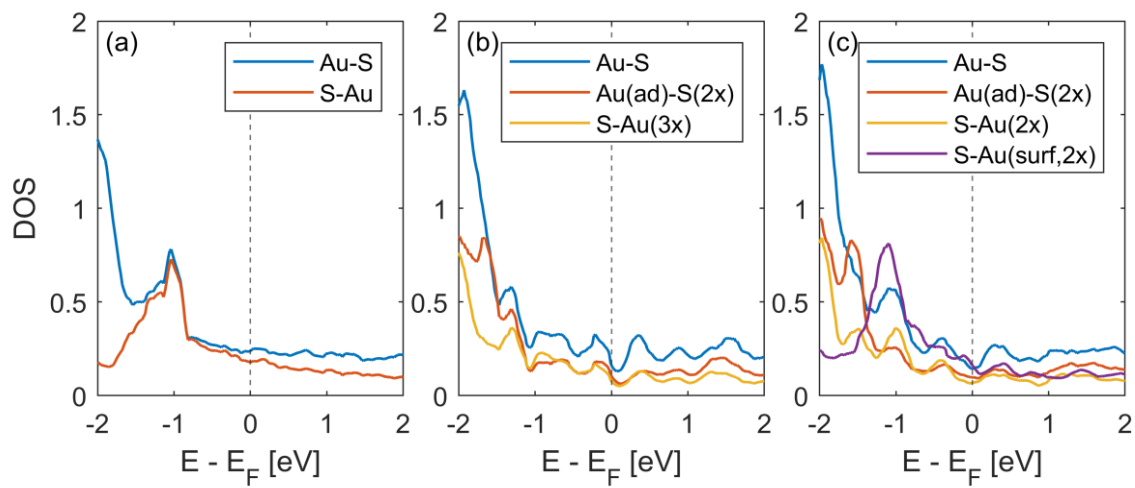

*Figure SI-8:* *Projected density of states relative to Fermi energy for methanethiol adsorbates, $SCH_3$, at a) bridge adsorption sites with 0.25 monolayer coverage on flat Au(111) and on b) Au-adatom-covered Au(111) with 1/3 and c) 3/8 monolayer thiol coverage. Only the DOSs projected on S atoms and Au atoms with Au-S bonds for each absorption structure are shown.*

Overall, most of the adsorption models studied here show a qualitatively similar behavior with the hybrid Au-S state at around -1 eV below at the Fermi energy in agreement with the expectations from the experiments. For the adsorption of atomic sulfur and at adatoms those peaks are shifted to more negative values, fitting less well to the experiments. Since the rest group and adsorption site for the other models do not seem to influence the DOS strongly, at least regarding the position of the hybrid HOMO states which is important here, we are confident to use those simplified models. Longer carbon chains could, however, lead to more pronounced steric effects, which might cause slight shifts in the DOSs. Furthermore, using adsorbates or even a mixture of molecules with, e.g., different headgroups or dipole moments, should allow for tuning the electronic structure to specific needs.